\documentclass{elsart1p}
\usepackage{graphicx}
\usepackage{amssymb}
\begin{document}
\begin{frontmatter}
%
%
%
%
%
\title{Di-muon measurements in CBM experiment at FAIR}
%
%

\author{A. Prakash$^1$, P. P. Bhaduri$^2$, S. Chattopadhyay$^2$, A. Dubey$^2$ $\&$ B. K. Singh$^1$}
\address{$^1$Department of Physics,Banaras Hindu University, Varanasi-221 005, India}\address{$^2$Variable Energy Cyclotron Centre, 1/AF, Bidhan Nagar, Kolkata-700 064, India}

\begin{abstract}
The compressed baryonic matter (CBM) experiment at the future FAIR accelerator facility near Darmstadt, Germany, aims at the investigation of baryonic matter at highest net baryon densities but moderate temperatures, by colliding heavy-ions at beam energies from 10 to 45 A GeV. The research program comprises the exploration of some basic landmarks of the QCD phase diagram like transitions from hadronic to partonic phase, the region of first order de-confinement as well as chiral phase transition, and the critical end point. The proposed key observables include the measurement of  low mass vector mesons and charmonia, which can be detected via their decay into the di-lepton channel. As the decayed leptons leave the hot and dense fireball without further interactions, hence they provide almost unscathed information about the interior of the collision zone where they are being created. In this paper, we discuss the physics motivation, detector concepts, and the feasibility studies in the di-muon measurements for central Au + Au collisions, with a special reference to the detailed simulation activities performed by the CBM muon group.We also discuss the R$\&$D activities of detector in brief. 

\end{abstract}

\begin{keyword}FAIR, CBM, QCD phase diagram, deconfinement, chiral symmetry, vector mesons, charm, muon, PLUTO, HSD, URQMD, GEANT3, GEM 
%
\PACS 25.75.-q
\end{keyword}
\end{frontmatter}

\section{Introduction}
\label{}

The exploration of the phase diagram of strongly interacting matter is one of the most intriguing fields of particle physics. One of the typical interests is the transition from hadronic to partonic matter at high temperature and/or high baryon densities. The experiments at RHIC and LHC are devoted to investigate the properties of deconfined QCD matter at very high temperatures and almost close to zero baryon densities. On other hand  the high luminosity CBM experiment at FAIR will investigate matters at high baryon density and moderate temperature. The research program comprises the study of the structure and the equation-of-state of baryonic matter at densities comparable to the ones in the inner core of neutron stars. This includes the search for the first order phase transition from hadronic to partonic matter, the critical endpoint and the search for signatures for the onset of chiral symmetry restoration on high net baryon densities.

One of the important aspects of the CBM experiment is to look for rare probes like charmonia (J/$\psi$, $\psi'$ etc.)  having extremely low production cross section in the FAIR energy regime. Charmonia as well as low mass vector mesons ($\rho$, $\omega$, $\phi$) can be measured via their decay in di-muons. The physics motivation of  di-muon measurement is to look for observables sensitive to the in-medium modifications of hadrons and the predicted first order phase transition from hadronic to quark-gluon matter at very high net baryon densities. Short-lived neutral vector mesons with di-lepton decay channels are very promising probes of in-medium effects. The deduced modification of the spectral function might be related to the expected restoration of chiral symmetry in the dense medium \cite{rapp}. No measurements have been  performed so far on dilepton production in heavy-ion collisions in the beam energy range between 2 to 35 AGeV. Thus dilepton data from CBM will be highly welcome. The production and propagation of charm in heavy-ion collisions is expected to be a particularly sensitive probe of the hot and dense medium. The 'anomalous suppression' in charmonium production (in addition to 'normal nuclear absorption' also present in p+A collisions), in heavy-ion collisions, has long been predicted as a 'smoking gun signature' for the formation of color deconfined medium \cite{SATZ}. No data on J/$\psi$ production are available in nucleus-nucleus collisions at beam energies below 158 AGeV. At FAIR charm production will be studied at beam energies close to the kinematic threshold and the production mechanisms of charmonium are expected to be sensitive to the conditions inside the early fireball

\section{CBM Detector Concept}
The proposed  layout is shown in figure 1, with the setup for muon identification \cite{peter2}. It consists of a Silicon Tracking System (STS) inside a dipole magnet as the primary tracking device. The muon detection system which will track the particles after STS will consist of series of iron absorbers, for hadron absorption and a number of tracking detectors sandwiched between them. The present optimized design includes 6 iron absorbers and 18 detector layers (3 behind each absorber). The total absorber length in the current design amounts to 2.25 m of iron. An additional shielding is used around the beam pipe in order to reduce the background of secondary muons produced in the beam pipe.

\begin{figure}
\centering
\includegraphics[width=8 cm]{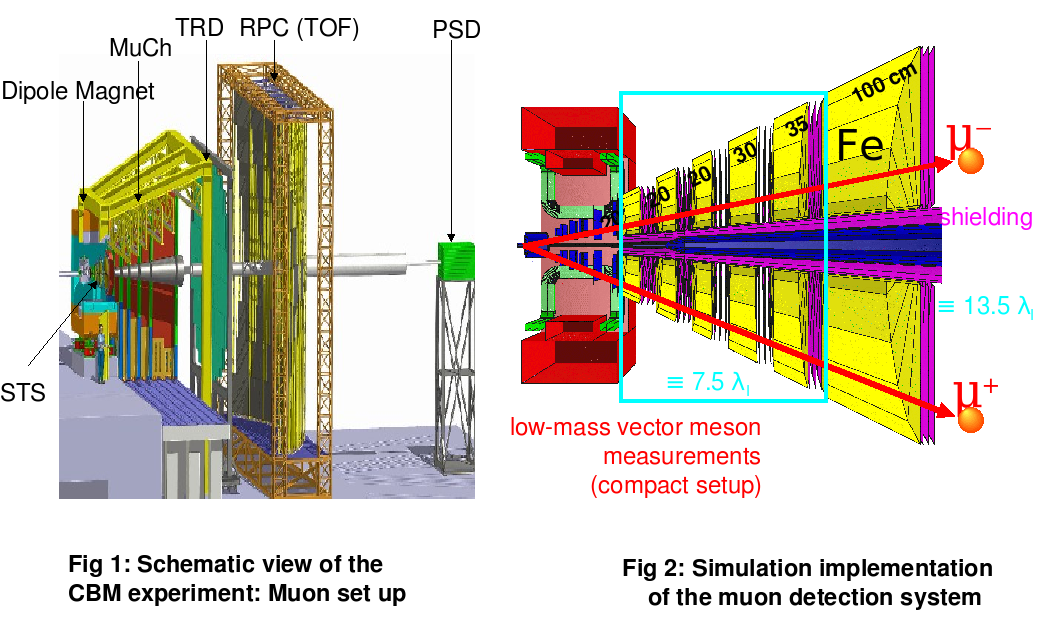}
\end{figure}

\section{Simulation procedure and results}

Simulations are being performed for the optimization of the detector design and to study the feasibility of the di-muon measurement. The feasibility studies are done within the CBM simulation framework \cite{cbmroot} which allows full event simulation and reconstruction.
 The ingredients used for the simulation are :
 a) PLUTO \cite{Pluto} generator for phase space decay of the vector mesons taking multiplicities from HSD \cite{HSD} 
 b) URQMD \cite{urqmd} generator for background particles and 
 c) GEANT3 \cite{geant} for transport of the generated particles through the setup.
d) Kalman Fitter (KF) for tracking

 The detection procedure involves the reconstruction of the track parameters in STS and extrapolation to muon detecting stations through the absorbers. Selection of the number of muon stations decide the identification of muons from low mass vector mesons (LMVM) and charmonia. While the LMVM muons travel shorter distances, J/$\psi$ muons cross the thick absorber and reach till the end. We have therefore taken tracks travelling through 15 layers and 18 layers as valid muon candidates from LMVM and charmonia respectively. We have segmented the detector into pads of varying size from 4mm$\times$4mm to 3.2cm$\times$ 3.2 cm depending upon the radial distribution of particle density. The reconstruction efficiency and signal to background ratio (S/B) of $\omega$ $\& $J/$\psi$ mesons were calculated in  a $\pm$2$\sigma$ window around the signal invariant mass peak and are presented in Table 1, for central Au-Au collisions at 8,25 and 35 AGeV beam energies. Figures 3 $\&$ 4 show the invariant mass spectra of $\omega$ and J/$\psi$ via di-muon channel. The combinatorial background is calculated using Super Event (SE) Analysis technique where tracks having opposite charges from different UrQMD events are combined. Our studies indicate that both low mass vector mesons and charmonia can be identified above the combinatorial background which is dominated by muons from weak pion decays.

\begin{figure}
\centering
\includegraphics[width=7 cm]{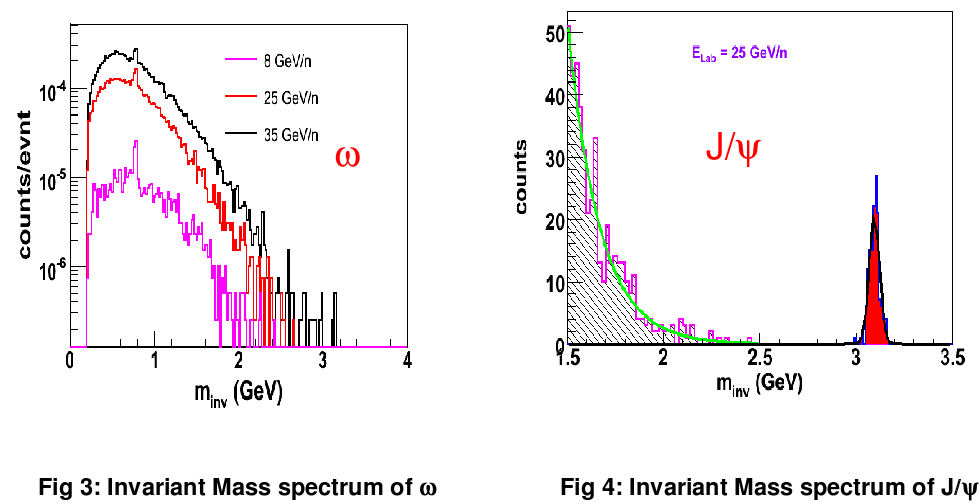}
\end{figure}

\section{R$\&$D on Detectors}

 Prototype detectors using 10 cm x 10 cm triple-GEM stack with a drift mesh and pad readout have been built at VECC. The chambers have been tested with radioactive sources and proton beams. Argon (Ar) and Carbon dioxide (C$O_2$) mixed in the ratio 70:30, was used as the gas mixture.
Cosmic tests performed with one such chamber yielded a charged particle detection efficiency of 95 \% . A set of chambers were tested with 2.5 GeV/c proton beams at SIS-18 beam line at GSI. Fig 6 shows the beam spots as seen on two chambers. The corresponding pulse height spectra is fitted with  Landau distribution functions for different operating voltages and the pulse height variation with high voltage is observed to be linear.

\begin{figure}
\centering
\includegraphics[width= 7 cm]{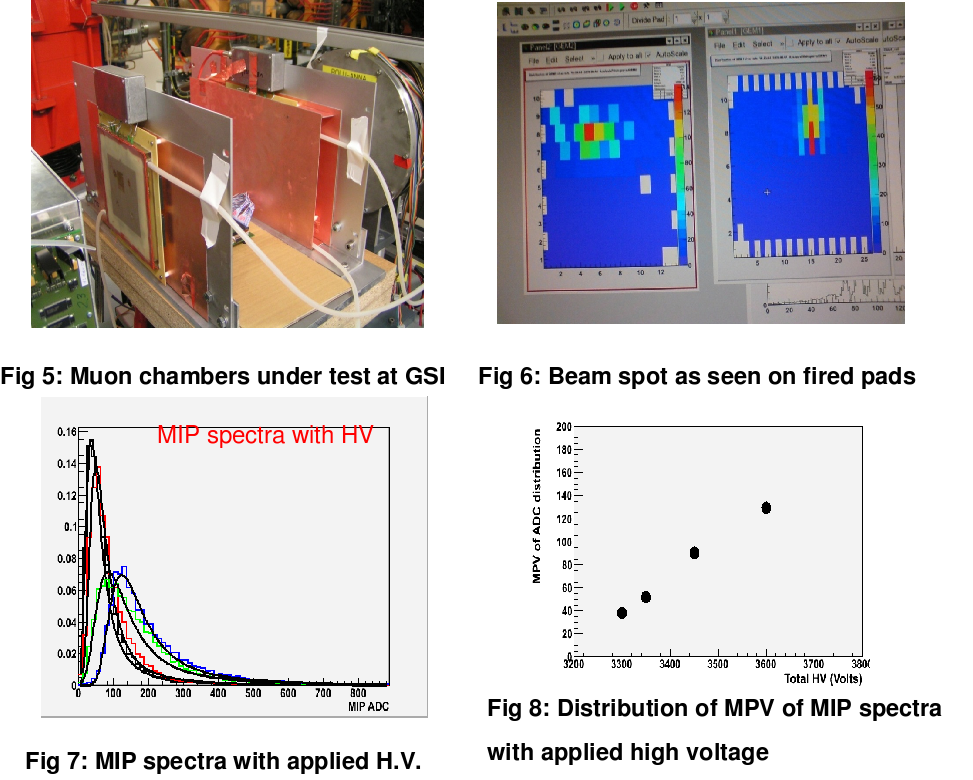}
\label{fig:5}
\end{figure}

\begin{table}
\centering
\caption{:Reconstruction Efficiency and (S/B) ratio of 
J/$\psi$ and $\omega$ in central Au-Au collision at 8,25 and 35 AGeV beam energies(Input events: 10k UrQMD+PLUTO)}

\begin{tabular}{|l|l|l|l|l|}
\hline 
\multicolumn{1}{|l|}{Energy (A GeV)}&\multicolumn{2}{l|}{Efficiency (\%)}&\multicolumn{2}{l|}{S/B} \\ \hline
\cline{2-5}
          &J/$\psi$$\rightarrow$$\mu^+$$\mu^-$  &$\omega$$\rightarrow$$\mu^+$$\mu^-$ &J$\psi$$\rightarrow$$\mu^+$$\mu^-$  &$\omega$$\rightarrow$$\mu^+$$\mu^-$\\ \hline
 8        & 4.9          & 0.96        & 3.3    &1.41 \\ \hline
25        & 13           & 1.58        & 7      &0.49  \\\hline
35        & 13           & 1.82        & 11     &0.34  \\\hline
\hline

\end{tabular}

\end{table}

\section{Summary and Outlook}
The CBM experiment at FAIR aims to explore the QCD phase diagram in the regime of  high densities and moderate temperatures. The high luminosity beam at CBM will help in detection of diagnostic probes with extremely low production cross-section (open and hidden charms, multi strange particles and other exotica). Muons are potential candidates for the measurement of observables like low-mass vector mesons and charmonia. A detector configuration with multi-layers of absorber-chamber combination has been optimized for detection of LMVM $\&$ charmonia at FAIR energy. A configuration of 6 absorbers of varying thickness and 18 detector layers having up to 1 million readout channels has been considered suitable for muon detection at FAIR.


\begin{thebibliography}{08}

\bibitem{rapp} H. van Hees, R. Rapp, arXiv:hep-ph/0604269v1
\bibitem{SATZ} Matsui T and Satz H 1986 Phys. Lett.B 178416
\bibitem{peter1} P. Senger, current proceeding
\bibitem{HSD} W. Cassing, E. Bratkovskaya, A. Sibirtsev, Nucl. Phys. A 786 (2007) 183-200
\bibitem{peter2} P. Senger J.Phys.G: Nucl. PArt. Phys. 31 (2005) S1111-S1114
\bibitem{Pluto} http://www$-$hades.gsi.de/computing/pluto
\bibitem{urqmd} S. A. Bass et al. , Prog. Part. Nucl. Phys. 41 (1998) 255.
\bibitem{geant} R. Brun et al., GEANT User Guide, 1986, CERN/DD/EE84-1.
\bibitem{cbmroot} http://cbmroot.gsi.de
\end{thebibliography}
\end{document}